\documentclass[reprint,superscriptaddress,amsmath,amssymb,aps,prl]{revtex4-2}

\usepackage[T1]{fontenc}
\usepackage{txfonts}
\usepackage{bm,dsfont,mathtools}
\usepackage{graphicx}
\usepackage{dcolumn}
\usepackage{tikz}
\usepackage{physics}
\usepackage{xcolor}
\usepackage{float}
\usepackage[colorlinks=true,breaklinks=true,allcolors=blue]{hyperref}

\usetikzlibrary{decorations.pathreplacing,decorations.markings,calligraphy}

\newcommand{\im}{\mathrm{i}}

\pgfdeclaredecoration{complete sines}{initial}
{
	\state{initial}[
	width=+0pt,
	next state=sine,
	persistent precomputation={\pgfmathsetmacro\matchinglength{
			\pgfdecoratedinputsegmentlength / int(\pgfdecoratedinputsegmentlength/\pgfdecorationsegmentlength)}
		\setlength{\pgfdecorationsegmentlength}{\matchinglength pt}
	}] {}
	\state{sine}[width=\pgfdecorationsegmentlength]{
		\pgfpathsine{\pgfpoint{0.25\pgfdecorationsegmentlength}{0.5\pgfdecorationsegmentamplitude}}
		\pgfpathcosine{\pgfpoint{0.25\pgfdecorationsegmentlength}{-0.5\pgfdecorationsegmentamplitude}}
		\pgfpathsine{\pgfpoint{0.25\pgfdecorationsegmentlength}{-0.5\pgfdecorationsegmentamplitude}}
		\pgfpathcosine{\pgfpoint{0.25\pgfdecorationsegmentlength}{0.5\pgfdecorationsegmentamplitude}}
	}
	\state{final}{}
}

\tikzset{
	photon/.style={decorate, draw=black,
		decoration={complete sines,amplitude=1ex, segment length=2ex}}
}

\makeatletter
\let\cat@comma@active\@empty
\makeatother

\begin{document}

\title{Prethermalization of light and matter in cavity-coupled Rydberg arrays}

\author{Aleksandr N. Mikheev}
\altaffiliation[current address: ]{Department of Physics, University of Konstanz, Universit{\"a}tsstra{\ss}e 10, 78464 Konstanz, Germany}
\affiliation{Institut f{\"u}r Physik,
		Johannes Gutenberg-Universit{\"a}t Mainz, 
		  Staudingerweg 7, 
		55128 Mainz, Germany}
\author{Hossein Hosseinabadi}
\affiliation{Institut f{\"u}r Physik,
		Johannes Gutenberg-Universit{\"a}t Mainz, 
		  Staudingerweg 7, 
		55128 Mainz, Germany}
\author{Jamir Marino}
\email{jamirmar@buffalo.edu}
\affiliation{Institut f{\"u}r Physik,
		Johannes Gutenberg-Universit{\"a}t Mainz, 
		  Staudingerweg 7, 
		55128 Mainz, Germany}
\affiliation{Department of Physics, 
        The State University of New York at Buffalo, 
        Buffalo, NY 14260, USA}

\begin{abstract}

We explore the dynamics of two-dimensional Rydberg atom arrays coupled to a single-mode optical cavity, employing nonequilibrium diagrammatic techniques to capture nonlinearities and fluctuations beyond mean-field theory. We discover a novel prethermalization regime driven by the interplay between short-range Rydberg interactions and long-range photon-mediated interactions. In this regime, matter and light equilibrate at distinct—and in some cases opposite—effective temperatures,   resembling the original concept of prethermalization from particle physics. Our results establish strongly correlated AMO platforms as  tools to investigate fundamental questions in statistical mechanics, including quantum thermalization in higher-dimensional systems.
\end{abstract}
\maketitle
\paragraph{Introduction} 
Finding mechanisms to avoid rapid thermalization in higher-dimensional quantum many-body systems remains a major open challenge in quantum statistical mechanics~\cite{Weimer2010,Kaufman2016,Choi2016,manovitz2024quantum}. Unlike their   one-dimensional counterparts, two- or three-dimensional quantum systems present significant conceptual and practical challenges. Central to these difficulties is the lack of a definition of quantum integrability beyond one dimension~\cite{Grabowski1995,Sutherland2004,Caux2011,Prosen2015}. The latter serves as a foundation   for understanding ergodicity through integrability-breaking ~\cite{DAlessio2016,Mitra2018,Polkovnikov2011}, which lies at the root of most of the experimental demonstrations of prethermalization~\cite{Le2023,Gring2012,Langen2015,Tang2018}.

Currently studied models, such as the quenched two-dimensional quantum Ising lattice, quickly thermalize, presenting a plain dynamical phase diagram~\cite{Fratus:2015,Blass:2016,Mondaini:2016,Mondaini:2017,Blass:2016,schmitt2020quantum,schmitt2022quantum,Hashizume2022,PhysRevLett.120.130601}. Notable distinctions are confinement~\cite{James2019} and the intrusion of domain walls~\cite{balducci2022localization,manovitz2024quantum}, which can trap the system into a long-lived nonthermal state.
While in one dimension prethermalization can arise simply from weak integrability breaking~\cite{Rigol2017,Kitagawa2011,Mitra2011,Marcuzzi2016,Kollar2011,Bertini2015,langen2016prethermalization,marcuzzi2013prethermalization},   the situation appears more intricate in higher dimensions. To date, explored scenarios include initial conditions with nontrivial topological structure ~\cite{Babadi:2015joa,PhysRevB.105.L060302,PhysRevB.108.144303,Rodriguez-Nieva2022},    kinetic constraints that restrict access to the full phase space~\cite{Feldmeier2019,PhysRevX.10.021041,signoles2021glassy,Valencia-Tortora2024,Schweizer2019,valencia2022kinetically}, and the emergence of nonthermal fixed points that induce self-similar scaling in the dynamics, thereby delaying thermalization \cite{Berges:2008wm,PineiroOrioli:2015cpb,Karl:2016wko,Deng:2018xsk,Spitz:2020,Heinen:2022ham,Heinen:2022rew,Noel:2025mtb}. Complementary to the latter,  quenching a system at the critical point (or across it) remains an ever-green option to induce prethermal dynamical scaling in the form of aging or coarsening~\cite{PhysRevB.88.024306,chiocchetta2017dynamical,PhysRevLett.115.245301,PhysRevB.94.174301,cugliandolo,bray1994theory}. The intense focus on prethermalization is   not only conceptually significant: avoiding rapid thermalization is essential for implementing meaningful many-body quantum information tasks.

In this work, we take a route inspired by strongly correlated AMO physics to achieve prethermalization of light and matter 
 in a platform   receiving increasing experimental attention   for its quantum processing role~\cite{Kong2021,Liu2023,Yan2023,Deist2022,Grinkemeyer2024,Hartung2024,hu2025site,puel2024confined,Bacciconi2025}. 
\begin{figure}
\centering
\includegraphics[width=\linewidth]{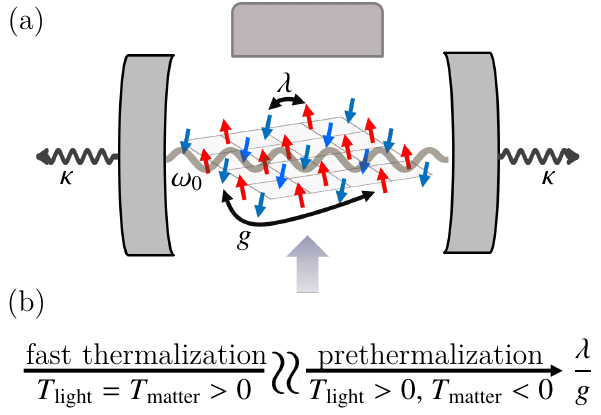}
\caption{(a) A two-dimensional Rydberg atomic array in a single-mode optical cavity.  Rydberg interactions $\lambda$ induce anti-ferromagnetism   between spins on neighboring sites, which competes against the long-range photon-mediated interaction of strength $g$. Photons leak from the cavity at rate $\kappa$. (b) Cartoon of the dynamical phase diagram. When photon-mediated long-range interaction plays the dominant role, the model displays fast thermalization, with matter and light quickly reaching the same temperature. In the opposite regime, light and matter prethermalize at different temperatures, featuring regimes where the atoms can stay  trapped in a metastable state characterized by a negative effective temperature.  
}
 \label{fig:setup}
\end{figure}
Specifically, we promote    Rydberg atom arrays embedded in optical cavities (see Fig.~\ref{fig:setup}a) as a paradigmatic model for exploring diverse thermalization dynamics in two dimensions. We demonstrate that the competition between short-range Rydberg interactions and photon-mediated long-range interactions leads to diverse dynamical responses. Depending on their relative strengths, the system either rapidly thermalizes or evolves into long-lived, nonthermal states.
Remarkably, in the Rydberg-dominated regime, we identify a novel prethermalization scenario where matter and radiation equilibrate independently at distinct temperatures, sometimes even reaching metastable states characterized by spins equilibrating at negative temperatures while the photon field remains at positive temperatures (see Fig.~\ref{fig:setup}b). Intriguingly, this appears close to the original concept of prethermalization in particle physics~\cite{Berges2004}, where distinct system components transiently equilibrate at separate temperatures before achieving global thermal equilibrium (for a related mechanism in one dimensional  quantum condensates, see Refs.~\cite{foini,agarwal2014chiral}).

This rich variety of dynamics emerges just through a straightforward modification of the paradigmatic two-dimensional quantum Ising model. The key difference is that the transverse field, instead of being fixed, is determined self-consistently by light-matter exchange in the form of the dynamical photon field coupled to the spins, showing that even the minimal inclusion of a photon in two-dimensional quantum spin lattices suffices to attain a rich set of prethermal responses. 

\paragraph{Model}

A two-dimensional square lattice of $N$ Rydberg atoms is situated inside a single-mode optical cavity \cite{Mivehvar2021,Kirton2019,Gelhausen:2016}, as depicted in Fig.~\ref{fig:setup}a. In the frame rotating with the laser pump frequency, the system's Hamiltonian can be written as
\begin{align}
\label{eq:H}
\hat{H} 
&=
\Delta\sum_{\mathbf{i}=1}^N \hat{s}_{\mathbf{i}}^z + \frac{\lambda}{4} \sum_{\langle \mathbf{i},\mathbf{j}\rangle}\left(\frac{\mathds{1}}{2} + \hat{s}^z\right)_\mathbf{i}\left(\frac{\mathds{1}}{2} + \hat{s}^z\right)_\mathbf{j}\nonumber\\
&+
\frac{2 g}{\sqrt{N}} \left(\hat{a} + \hat{a}^{\dagger}\right)\sum_{\mathbf{i}=1}^N \hat{s}_{\mathbf{i}}^x + \omega_0\hat{a}^{\dagger} \hat{a}\,.
\end{align}
Here, $\hat{s}_{\mathbf{i}}^{\alpha}$ and $\hat{a}^{\dagger}$ ($\hat{a}$) are the spin-$1/2$ operators defined on a lattice site $\mathbf{i}$ and the photon creation (annihilation) operators, respectively, satisfying the commutation relations $\left[\hat{s}_{\mathbf{i}}^{\alpha},\hat{s}_{\mathbf{j}}^{\beta}\right] = \im \delta_{\mathbf{ij}}\varepsilon^{\alpha\beta\gamma} \hat{s}^{\gamma}_{\mathbf{i}}$ and $[\hat{a},\hat{a}^{\dagger}] = 1$. The cavity and atomic frequencies, in this frame, are given by $\omega_0$ and $\Delta$, accordingly, with the latter playing the role of an effective longitudinal field in the spin language. The summation in $\sum_{\langle\mathbf{i},\mathbf{j}\rangle}$ goes over all the   neighboring lattice sites for each site $\mathbf{i}$. In addition, the system may experience single-photon losses at a rate $\kappa$, with the corresponding Lindblad jump operator $\hat{a}$ \cite{Breuer:2007juk}.

The cavity photons collectively couple to the atoms with the strength $g$, mediating a long-range interaction across the entire system. This term can be seen  as a self-consistent   transverse field $B^x = -2 g\langle\hat{a} + \hat{a}^{\dagger}\rangle/\sqrt{N}$ in a two-dimensional quantum Ising model  with the $\hat{s}^z_\mathbf{i}\hat{s}^z_\mathbf{j}$ interactions provided by the Rydberg coupling  $\lambda$, where we have taken the expectation value of the photon field in a first, crude approximation. 
This effect is crucial for generating nontrivial dynamics since Eq.~\eqref{eq:H} reduces to the classical Ising model for $g\to0$. In the remainder of the article, we adopt this heuristic picture to interpret the dynamics of the model, although it should be noted that this remains quantitatively  correct only on short time scales where correlations between light and matter are weak.
We also note that   light-matter interactions favor uniform spin configurations together with a nonzero expectation value of the photon operator~\cite{Kirton2019}, breaking the $\mathbb{Z}_2$ parity symmetry $\hat{s}^x \to -\hat{s}^x$, $\hat{a} \to -\hat{a}$ (superradiant phase).
This interaction competes with the local Ising-type antiferromagnetic interaction of strength $\lambda$ between spins on neighboring lattice sites. The latter energetically favors spin configurations with antiparallel $z$-components on adjacent sites, which break the lattice translation invariance.  The resulting equilibrium phase diagram has been explored in  \cite{Gelhausen:2016,puel2024confined,Schellenberger:2024cxk,Rohn2020,An:2022xmd}  for simple bipartite lattices and in \cite{Liang:2025ofd} for a triangular lattice.

We investigate equilibration dynamics of the system by solving a set of self-consistent and conserving (Kadanoff--Baym) equations  \cite{Danielewicz:1984,Berges:2004yj}, allowing us to capture nonlinearities and fluctuations beyond the reach of approximations based on cumulant expansions~\cite{Vardi:2001,Koehler:2002,Kramer:2015,Kirton:2018,Sanchez-Barquilla:2020,Robicheaux:2021,Plankensteiner:2022,Rubies-Bigorda:2023,Fowler-Wright:2023} and semiclassical methods~\cite{Schachenmayer2015,Mink2022,Wurtz2018,Hosseinabadi2025,Polkovnikov2010}, including at the same time effects responsible for thermalization~\cite{Babadi:2015joa,Batini:2024,Schuckert:2018fwg,Hosseinabadi:2023cyj,Hosseinabadi2024,PhysRevB.96.014512}. In order to handle path integrals for spins, we represent spin degrees of freedom on each lattice site as bilinears of Majorana fermions \cite{Berezin:1977,Tsvelik:1992,Coleman:1993,Shnirman:2003,Schad:2014,Babadi:2015joa,Hosseinabadi:2023cyj} (see also \cite{SM} for a comprehensive survey of the methodologies used). 
We consider initial states in the  tensor product of light and matter components,  $\ket{\psi_0} = \ket{\varphi_A,\theta_A}\otimes\ket{\varphi_B,\theta_B}\otimes\ket{0}$, where $\ket{0}$ denotes the photon vacuum state, while $\ket{\varphi,\theta}$ is the spin coherent state, parametrized with the azimuthal angle $\varphi$ and the polar angle $\theta$. To allow for antiferromagnetic order without breaking spatial homogeneity, we additionally split the original lattice into two magnetic sublattices, denoted by the subscripts $A$ and $B$ above. Our focus is to analyze how competing long- and short-range interactions can dynamically build correlations starting from the  classical  states $\ket{\psi_0}$. Note also that depending on the values of the angles, the energy of the initial state with respect to the post-quench Hamiltonian can be too high~\footnote{Quenching     here      means   the explicit initialization of the system in a nonequilibrium state -- distinct from the ground state of the Hamiltonian governing its subsequent evolution -- rather than  a sudden change in the Hamiltonian parameters.} or too close to the ground-state energy. In the former case, much of the dynamics could become trivial since order would  melt both transiently and at long times, while in the latter case, the dynamics could become extremely slow~\cite{Polkovnikov2011,Mitra2018,Essler:2016ufo}. For this reason, in the following we have adopted values of $\theta_{A,B}$ and $\varphi_{A,B}$ that avoid these two regimes, and therefore render  the initial condition both experimental-friendly and generic from the point of view of quench dynamics.

\begin{figure}[!t]
\centering
\includegraphics[width=0.95\linewidth]{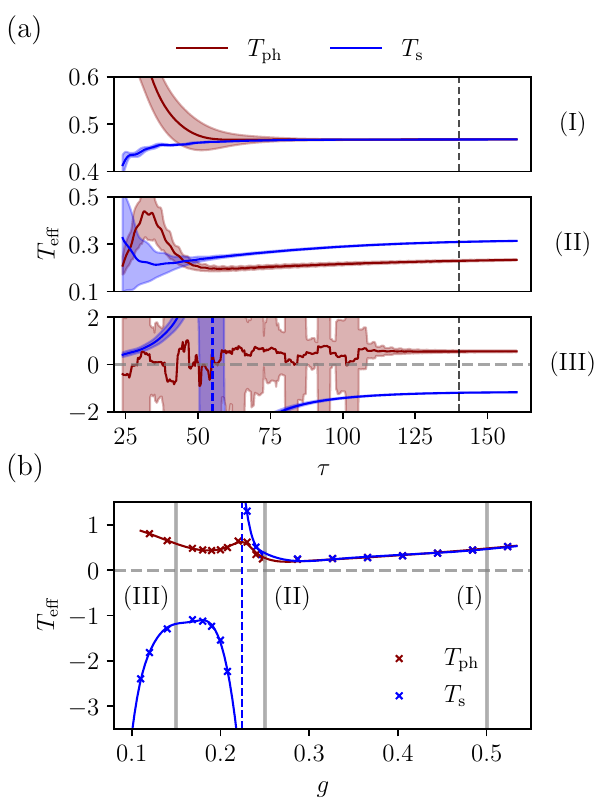}
\caption{ 
(a) Time dependence of the spin and photon effective temperatures. The parameters are taken as $(\Delta,\lambda) = (-0.1,0.5)$ and $g = 0.5, 0.25$, and $0.15$ for regimes I, II, and III, respectively. The semitransparent ribbons represent the uncertainty arising from averaging $T_{\mathrm{eff}}$ over small frequency and time windows according to the procedure detailed in  \cite{SM}. Initially large,   temperature deviations gradually vanish, signaling a transition to the regime where the notion of effective temperatures becomes reliable. (b) Spin and photon effective time temperatures   taken at time $\tau=140$, and  displayed as a function of the Dicke coupling $g$ (we mark this time in panel (a) by the vertical black dashed line). The gray vertical lines indicate the values of $g$ used in the respective regimes in (a).}
 \label{fig:Ts_ph}
\end{figure}

\paragraph{Correlation functions and effective temperature} 
We explore the dynamics of the model by analyzing statistical correlations and spectral properties of the atomic and photon degrees of freedom. 
For bosons, the spectral function $\rho$ corresponds to the expectation value of the commutator of the creation/annihilation operators, while the statistical function $F$ is given by the connected part of the expectation value of the anti-commutator. 
The former ($\rho$) encodes information on the spectral nature of single or collective excitations in the system, while the symmetric correlation function ($F$) informs about the statistical distribution (Gibbs or nonequilibrium) of these excitations~\cite{kamenev2023field}. In thermal equilibrium, they  combine to form a fluctuation-dissipation relation~\cite{altland2010condensed} 
\begin{equation}
\label{eq:FDR}
F_{\mathrm{eq}}(\omega) = \left[n_{\mathrm{eq}}(\omega) + \frac{1}{2}\right]\rho_{\mathrm{eq}}(\omega) =  \frac{1}{2} \coth{(\omega/2T)}\,\rho_{\mathrm{eq}}(\omega)\,,
\end{equation}
which naturally motivates to define the effective temperature $T_{\mathrm{eff}}(\tau,\omega) = \omega/\lbrace 2\mathrm{acoth}\left[2 n(\tau,\omega) + 1\right]\rbrace$ out of equilibrium by promoting   $n_{\mathrm{eq}}(\omega) \to n(\tau,\omega) \equiv F(\tau,\omega)/\rho(\tau,\omega) - 1/2$, $\rho_{\mathrm{eq}}(\omega) \to \rho(\tau,\omega)$ \cite{Aarts:2002,Arrizabalaga:2005,Scarlatella:2019}. Here, $O(\tau,\omega)$ denotes the Fourier transform of $O(t,t')$ with respect to the relative coordinate $s=t-t'$, also known as the Wigner transform \cite{Berges:2004yj}, while dynamics is in central time $\tau = (t+t')/2$. 

In literature, the above definition of $T_{\mathrm{eff}}$ is often applied to nonequilibrium steady states \cite{Marquardt:2007,Clerk:2010,Pagel:2013,Babadi:2015joa,Levitan:2016,Hsiang:2020,Young:2020,hosseinabadi_YSYK2023}, in which case the effective temperature has a direct operational meaning: by probing the system via a qubit with frequency $\omega_{\mathrm{probe}}$, the latter will thermalize to a temperature $T_{\mathrm{eff}}(\omega\simeq\omega_{\mathrm{probe}})$. However, the notion of a nonequilibrium effective temperature can be useful even for certain time-evolving states. Since the model~\eqref{eq:H} is non-integrable, a generic initial state is expected to effectively thermalize such that ${n(\tau\to\infty,\omega) \to n_{\mathrm{eq}}(\omega)}$, with lower-frequency modes tending to equilibrate faster than higher-frequency modes, which may still exhibit strong nonthermal features even at late times ~\cite{Strack:2013,Maghrevi:2016,Young:2020,Sieberer:2023}. This motivates us introducing an effective low-frequency temperature to characterize the intermediate regime of the thermalization dynamics~\cite{Aarts:2002,Arrizabalaga:2005}. During this regime  the infrared region of the frequency spectrum has already relaxed, while the system continues to slowly evolve toward full equilibrium at higher energies. Importantly, the above definitions allow us to independently extract the effective temperatures for the photonic $T_{\mathrm{ph}}$ and spin $T_{\mathrm{s}}$ degrees of freedom, respectively. In this work, the effective temperature for spin degrees of freedom is defined with respect to local spin-$z$ correlation functions. For more details on the numerical extraction procedure of $T_{\mathrm{ph}}$ and $T_{\mathrm{s}}$, we refer to  \cite{SM}.

\begin{figure}[t]
\centering
\includegraphics[width=\linewidth]{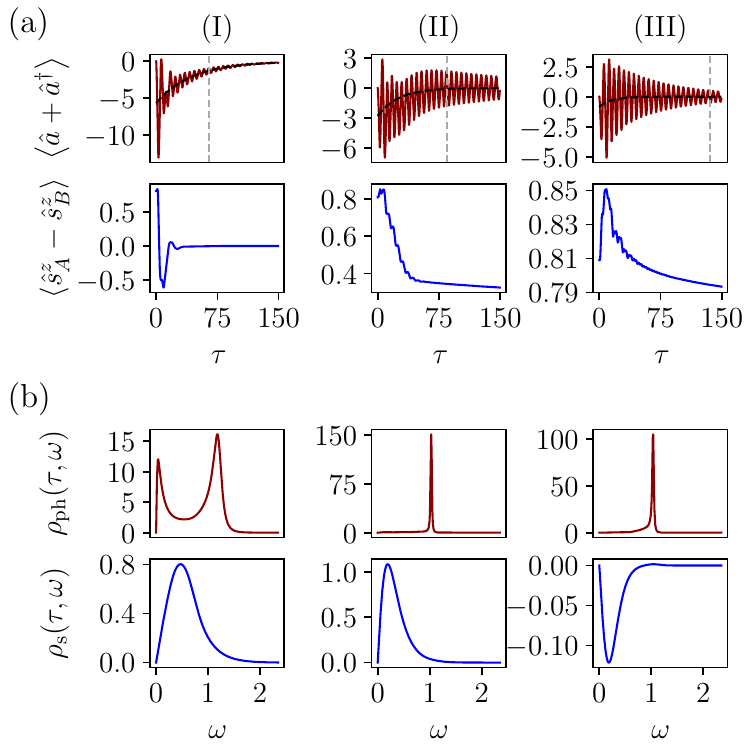}
\caption{Plots of observables for the same initial conditions and values of coupling constants as in Fig.~\ref{fig:Ts_ph}. (a) Time dependence of the   photon coherence (top) and of the staggered magnetization (bottom).   Gray dashed lines mark the onset of (pre)thermalization as extracted from Fig.~\ref{fig:Ts_ph}. (b) Photon (top) and spin (bottom) spectral functions at time $\tau=140$ across the three dynamical regimes. In regime III, the spin spectral function is inverted at low frequencies, signaling a negative effective temperature. }
\label{fig:observables}
\end{figure}

\paragraph{Thermalization dynamics}
In Fig.~\ref{fig:Ts_ph}a, we show the time dependence of the low-frequency effective temperatures for three choices of parameters  starting from the same initial condition $\ket{\psi_0} = \ket{0,0.2\pi}\otimes\ket{0,0.8\pi}\otimes\ket{0}$. The system is taken on a two-dimensional square lattice, with $10\times10$ spins on each of the two sublattices. All quantities presented in the following are assumed to be measured in units of $\omega_0$, which we accordingly set to one. The remaining parameters are taken as $(\kappa,\Delta,\lambda) = (0.0,-0.1,0.5)$ and $g = 0.5, 0.25$, and $0.15$ for three regimes, labeled I, II, and III. The effect of photon losses is negligible for times $\tau \lesssim 1/\kappa$, as we have confirmed numerically. As a consistency check, we have benchmarked our simulations with dynamics truncated at the semiclassical level~\cite{Hosseinabadi2025,Polkovnikov2010}, and found good agreement on short time scales. Such behavior is expected since, after a short transient period, the system becomes strongly correlated and the subsequent dynamics can be reliably described only within our field-theoretic approach.

Regime I is characterized by strong long-range interactions; the system demonstrates fast thermalization, with matter and light quickly reaching the same temperature. From the Hamiltonian perspective, this regime can be considered as a perturbed Dicke model in the superradiant phase, which is known to show fast effective thermalization \cite{Altland:2012,Lewis_Swan:2019,Villasenor:2022}. According to Fig.~\ref{fig:observables}a, the photon coherence decays with time in regime I. This decay arises from finite-size effects~\cite{Hosseinabadi:2023cyj} and fluctuations induced by the Rydberg interactions. Notably, it does not imply the destruction of superradiance.  The presence of a superradiant state is manifested by the scaling of photon two-point functions with the system size, $\langle a^2 \rangle \sim N$, see \cite{SM} for more details. A further hallmark of superradiance is the critical spectrum of photons (upper left panel of Fig.~\ref{fig:observables}b), with a pronounced peak at $\omega \to 0$ due to strong hybridization and entanglement between spins and photons~\cite{Strack:2013,Kirton2019,Emary_2003}.

 In spin language, the photon coherence can be viewed as an effective dynamical transverse field ${B^x(t) \sim -\langle\hat{a} + \hat{a}^{\dagger}\rangle}$.  For $B^x=0$, the resulting effective Hamiltonian is diagonalizable in the local basis given by the eigenstates of $\hat{s}^z_{\mathbf{i}}$, implying absence of thermalization. In contrast, when this self-consistent field is nonzero  for a sufficiently long time window, cf. Fig.~\ref{fig:observables}a (upper left panel), we get an effective two-dimensional transverse-field quantum Ising model which quickly thermalizes ~\cite{schmitt2020quantum,Blass:2016,PhysRevLett.120.130601}.

Quenching the dynamics with smaller values of the long-range coupling, we reach a point where the system does not realize a superradiant phase anymore. 
The absence of superradiance can be inferred both from the evolution of the photon condensate, which decays while showing strong oscillations (upper middle panel of Fig.~\ref{fig:observables}a), and more importantly, from the trivial photonic spectrum featuring a narrow line at the cavity frequency at late times (upper middle panel of Fig.~\ref{fig:observables}b), which has to be contrasted with the broad photonic spectrum in the superradiant phase (upper left panel of Fig.~\ref{fig:observables}b). While the ground state of the post-quench Hamiltonian is superradiant, the   critical temperature   is nearly an order of magnitude smaller in this case compared to regime I, and the order melts due to the excess energy injected by the quench. Up to intermediate times the effective field $B^x$ is dominated by strong temporal fluctuations  (upper middle panel of Fig.~\ref{fig:observables}a) that act as a fast noisy drive which heats up the atoms, {cf. also Ref.~\cite{Kozlej:2025omn} for a similar discussion}.
The weak long-range coupling allows the stabilization of antiferromagnetic order in the system supported by $\lambda$, as can be seen from the behavior of staggered magnetization which saturates to a finite value (lower middle panel of Fig.~\ref{fig:observables}a). However, the most notable feature   is the emergence of a long-lasting prethermal state in which spins and photons coexist at different temperatures (middle panel of Fig.~\ref{fig:Ts_ph}b). The presence of this state, which is in stark contrast with behavior of generic interacting systems~\cite{Polkovnikov2011,DAlessio2016,Mitra2018}, can be attributed to the lack of proper energy resonances between the spin and photon sectors (after the initial fast energy transfer   enabled by $B^x(t)$). At such late times  (dashed lines in Fig.~\ref{fig:Ts_ph}a), the energy exchange rate between spins and photons is, according to the Fermi’s golden rule, proportional to the overlap of spin and photon spectral densities as $\mathrm{d}E/\mathrm{d}t \propto \int \dd{\omega} \omega\, [\ldots]\,\rho_\mathrm{ph}(\omega)\,\rho_s(\omega)$, where $[\ldots]$ represents inconsequential  spin and photon occupation functions. Considering the small overlap of these two in regime II, as compared to regime I (cf. Fig.~\ref{fig:observables}b left and central panels), we expect inefficient energy transfer between spins and photons at late times, in agreement with the observed prethermal plateaus. The spectral argument is further supported by our observation that upon adding photonic losses, the photon spectral line broadens and the prethermal state disappears at long times with all subsystems reaching the same temperature. 

Modern c-QED platforms, however, allow for cavity loss rates that are   orders of magnitude smaller than the cavity frequency $\omega_0$~\cite{Mivehvar2021}, suggesting that the proposed prethermalization phenomenon can be experimentally observed, up to times $t\lesssim \omega^2_0/g^2 \kappa$ (see for instance Ref.~\cite{Agarwal:1997}).  Furthermore, one of the key advantages of Rydberg atoms in AMO platforms   is their exceptionally long lifetimes relative to spontaneous emission. In contrast to typical optical transitions, Rydberg states can exhibit decay rates that are up to five orders of magnitude slower~\cite{saffman2010quantum,browaeys2020many}. At the same time, the atomic detuning $\Delta$ can serve   to control the boundaries between the normal phase and the superradiant one. The critical atom-light coupling scales in this case as $g_c\sim\sqrt{|\Delta|}$~\cite{Gelhausen:2016}, allowing to push the prethermalization window to larger values by controlling the strength of the Dicke coupling $g$, which is an established feature in c-QED experiments~\cite{Mivehvar2021,Baden:2014}. Likewise, the strength of Rydberg interactions is highly tunable, ranging from small values up to the deep blockade regime,     $ \lambda/2\pi \approx 500\,\mathrm{MHz} $~\cite{Semeghini:2021wls}.

\paragraph{Prethermalization at negative temperatures}

Finally,    when Rydberg interactions are dominant, regime II crosses over to  regime III, with persisting   prethermalization of light and matter. Once   again this can be attributed to the absence of resonances between the spin and photon subsystems at late times (right panel of Fig.~\ref{fig:observables}b). The distinct feature of regime III, however, is the emergence of a negative temperature spin state coexisting with positive temperature photons. Usually, a negative temperature indicates an inverted state in systems with a bounded energy spectrum, with a total energy close to the upper edge of the spectrum. This observation is highlighted by the late-time spin spectral function switching signs, as shown in Fig.~\ref{fig:observables}b (right panel). At thermal equilibrium, the spin spectral function follows from $\rho_s(\omega) \propto \sum_{ij} (p_j-p_i)\,\abs{\bra{i}\hat{s}^z\ket{j}}^2 \delta(\omega - E_i + E_j)$, and the property $\omega \rho_s(\omega)>0$ can be traced back to the condition that lower-energy states have a larger population than higher-energy states, $p_j \leq p_i$ whenever $E_j \geq E_i$.  A violation of this condition indicates the existence of population inversion, a hallmark of negative-temperature steady states \cite{Levitan:2016,Scarlatella:2019,Motazedifard:2023}.

The inverted state is formed due to the excessive energy injected into the spin sector during the quench. This can be understood by observing the steady-state behavior of $T_s$ in terms of $g$. Starting from regime III and approaching regime I, the spin temperature approaches $T_s \to + \infty$ on the right side of the transition (marked by the vertical dashed line in Fig.~\ref{fig:Ts_ph}b), followed by a sharp drop to $T_s \to - \infty$ on the left side, before reaching finite negative values in regime III. This behavior is strikingly similar to the dependence of a single spin’s temperature on its average energy~\cite{kittel}, which is given by $T(\epsilon)=\delta/\mathrm{arctanh}(\epsilon/\delta)$, where $2\delta$ is the energy splitting of the spin. The temperature approaches $T\to \pm \infty$ when $\epsilon \to 0 ^\mp$, and becomes negative for $\epsilon >0$. Accordingly, regime III describes a spin ensemble which is trapped in a highly excited state due to the lack of efficient mechanisms to lose its energy. Under realistic circumstances, the cavity loss and the spontaneous atomic decay are finite and eventually destroy the inverted state. However, compared to other energy scales in the system, both of these energy scales can be made typically small for high-finesse cavities and Rydberg-dressed atoms~\cite{Kong2021,Liu2023,Yan2023,Deist2022,Grinkemeyer2024,Hartung2024,hu2025site}.

\paragraph{Perspectives} 
 
We have demonstrated that the two-dimensional quantum Ising model  can exhibit prethermalization for a broad range of     parameters when its transverse field becomes dynamic and is self-consistently determined by light-matter interactions. In particular, light and matter can equilibrate at distinctly different temperatures, and notably, matter may sustain negative temperatures over extended periods of time. Our prethermalization is reminiscent of  its original formulation in   high-energy physics~\cite{Berges2004}, where distinct system components transiently equilibrate at separate temperatures before achieving global thermal equilibrium. 
 
At the same time, our findings put forward Rydberg arrays   in optical cavities as a promising new class of quantum simulators for spin models, characterized by the unique competition of short- and long-range interactions. For instance, we have recently investigated novel forms of quantum scarring~\cite{Hosseinabadi:2025xup} and have a work in preparation exploring topological spin liquids in these systems~\cite{darrick}, which would clearly show different dynamics than the one discussed in this article due to geometric frustration from the underlying lattice~\cite{Liang:2025ofd} (see also \footnote{We also expect dimensionality to play a role. For instance, in one dimension dynamics is known to be slower and this could significantly affect our findings.}). Together with advancements in other cavity-QED setups capable of surpassing traditional Dicke-type all-to-all connectivity~\cite{kelly2021effect,zhang2021observation,helson2023density,kroeze2023replica,periwal2021programmable}, our work heralds the entry of AMO physics into its strongly correlated era. \\

\paragraph{Acknowledgments} 
This project has been supported by the Deutsche Forschungsgemeinschaft (DFG, German Research Foundation): through Project-ID 429529648, TRR 306 QuCoLiMa (``Quantum Cooperativity of Light and Matter'')  and by the QuantERA II Programme that has received funding from the European Union’s Horizon 2020 research and innovation programme under Grant Agreement No 101017733 ``QuSiED'') and by the DFG (project number 499037529); and by the Dynamics and Topology Center funded by the State of Rhineland Palatinate.  This research was supported in part by grant NSF PHY-2309135 to the Kavli Institute for Theoretical Physics (KITP). We gratefully acknowledge the computing time granted through the project ``DysQCorr'' on the Mogon II supercomputer of the Johannes Gutenberg University Mainz (\href{https://hpc.uni-mainz.de}{hpc.uni-mainz.de}), which is a member of the AHRP (Alliance for High Performance Computing in Rhineland Palatinate, \href{www.ahrp.info}{www.ahrp.info}), and the Gauss Alliance e.V.

We  thank  S. Diehl, R.J. Valencia-Tortora, O. Zilberberg for helpful discussions;   Z. Bacciconi, F. Balducci, J. Zeiher for their insigthful comments on the manuscript; and   A. Rosch for inspiring conversations, which have initiated this project. 

\paragraph{Data availability} The data that support the findings of this work are openly available \cite{Zenodo}.

\begin{widetext}
	\begin{center}
		\textbf{Supplementary Material}
	\end{center}
	\renewcommand{\theequation}{S.\arabic{equation}}
	\renewcommand{\thefigure}{S\arabic{figure}}
	\setcounter{equation}{0}
	\setcounter{figure}{0}
	
In the following, we provide further details of the theoretical framework employed in this work, numerical methodology, and results.

\section{Synopsis   of the method }

We first provide a short summary of the technical steps employed for the benefit of the reader interested in skipping the details of the calculations.

We investigate the equilibration dynamics of the system by solving a set of self-consistent (Kadanoff--Baym) equations for the one- and two-point correlation functions on a Schwinger--Keldysh closed-time-path contour \cite{Danielewicz:1984,Berges:2004yj}. The contour two-point functions give us direct access to the spectral and statistical properties of the system, forming the core of our analysis. To that end, we first map the spin degrees of freedom on each lattice site to a set of Majorana fermions \cite{Berezin:1977,Tsvelik:1992,Coleman:1993,Shnirman:2003},
\begin{equation}
	\label{eq:majoranas}
	\hat{s}_{\mathbf{i}}^{\alpha} = -\frac{\im}{2}\left(\hat{\bm{\eta}}_{\mathbf{i}}\times\hat{\bm{\eta}}_{\mathbf{i}}\right)^{\alpha}, \quad \left\lbrace\hat{\eta}_{\mathbf{i}}^{\alpha},\hat{\eta}_{\mathbf{j}}^{\beta}\right\rbrace = \delta_{\mathbf{ij}}\delta^{\alpha\beta}\,,
\end{equation}
see also Refs.~\cite{Schad:2014,Babadi:2015joa,Hosseinabadi:2023cyj} for applications out of equilibrium. 

The integro-differential nonequilibrium Dyson equations \cite{Berges:2004yj} that govern the dynamics of the two-time two-point correlation functions $D(t,t')$ and $G(t,t')$ for the photon and Majorana degrees of freedom, respectively, consist of two parts. The first (differential) part, which can be schematically written as $\partial_t D + f(D,G)$, describes Markovian dynamics, similar to that obtained from the standard cumulant expansion approximation \cite{Vardi:2001,Koehler:2002,Kramer:2015,Kirton:2018,Sanchez-Barquilla:2020,Robicheaux:2021,Plankensteiner:2022,Rubies-Bigorda:2023,Fowler-Wright:2023}. The second (integral) part is represented by the convolution $\int\dd{t''} \Sigma(t,t'')\,D(t'',t')$ of the two-point correlator with the self-energy function $\Sigma$, and likewise for the Majorana degrees of freedom. This term breaks the Markovian nature of the approximation, introducing memory effects~\cite{babadi2013non}. The emergence of such memory effects is a natural consequence of a self-consistent reduction of the infinitely-dimensional state space, characterized by correlation functions of all orders, forming the BBGKY or, more generally, the Martin--Schwinger hierarchy \cite{Stefanucci:2013}, to a simpler description comprising only one- and two-point functions. The memory integral, which stores the system's interaction history, contains information about higher-order correlations, allowing to reconstruct the latter from the full history of the one- and two-point correlation functions. We use this property to reconstruct spin-spin correlators, which represent four-point functions in the language of Majorana fermions, cf. Eq.~\eqref{eq:majoranas}.  We stress that it is exactly the non-Markovian memory term that is responsible for the build-up of many-body correlations during, e.g., quasiparticle scattering processes, making it a key ingredient for describing thermalization dynamics.  

The self-energies encode the information about the structure of correlations induced by the nonlinearities and diagrammatically represent an infinite sum of one-particle irreducible diagrams with two amputated legs \cite{Weinberg:1995,Berges:2004yj}. The choice of diagrams used to approximate this infinite series is not unique and usually consists of a systematic expansion in some small parameter, the most popular choice of which is the interaction coupling constant. 
To go beyond the weak-coupling limit, we use instead a pair of nonperturbative expansion parameters. The first parameter, $1/N$, reflects the collective nature of the long-range interaction term, suppressing fluctuations as the system size increases. The fluctuations induced by the short-range antiferromagnetic interaction, however, are not affected by the system size and thus not controlled by the parameter $1/N$. To overcome this problem, we first construct an auxiliary model, in which the spin at each lattice site is replicated $N_s$ times, effectively increasing the spin length to $N_s/2$. After carrying out the nonperturbative $1/N_s$ expansion at next-to-leading order, we take the limit $N_s\to1$ corresponding to the original model \cite{Babadi:2015joa,Batini:2024}. These equations are supplemented with the equation for the expectation value of the photon operator $\langle\hat{a}\rangle$ and the initial conditions. 

In the following, we substantiate each of these technical steps.

\section{2PI formalism}
The Keldysh action corresponding to the Hamilonian \eqref{eq:H}, including potential single-photon losses, can be written as $S = S_0 + S_{\mathrm{int}} + S_{\mathrm{dis}}$, with
\begin{align}
	\label{eq:S}
	S_0
	&=
	\frac{\im}{2}
	\int_{\mathcal{C}}\dd{t} 
	\left[
	\left(a^{*}, a\right) 
	\begin{pmatrix}
		\partial_t + \im \omega_0 & 0 \\
		0 & -\partial_t + \im\omega_0
	\end{pmatrix}
	\begin{pmatrix}
		a\\
		a^{*}
	\end{pmatrix}
	+
	\left(\eta_{\mathbf{i},\sigma}^x, \eta_{\mathbf{i},\sigma}^y, \eta_{\mathbf{i},\sigma}^z\right) \begin{pmatrix}
		\partial_t & \Delta + \lambda & 0\\
		-\Delta - \lambda & \partial_t & 0 \\
		0 & 0 & \partial_t
	\end{pmatrix}
	\begin{pmatrix}
		\eta_{\mathbf{i},\sigma}^x\\
		\eta_{\mathbf{i},\sigma}^y\\
		\eta_{\mathbf{i},\sigma}^z
	\end{pmatrix}
	\right],\nonumber\\
	S_{\mathrm{int}}
	&=
	\int_{\mathcal{C}}\dd{t} \left[
	\frac{2 \im g}{\sqrt{N N_s}} \left(a + a^{*}\right) \eta_{\mathbf{i},\sigma}^y \eta_{\mathbf{i},\sigma}^z
	+
	\frac{\lambda}{4 N_s} T_{\mathbf{ij}} \eta_{\mathbf{i},\sigma}^x\eta_{\mathbf{i},\sigma}^y\eta_{\mathbf{j},\sigma'}^x\eta_{\mathbf{j},\sigma'}^y\right],\quad
	S_{\mathrm{dis}}
	=
	- \im \kappa \int\dd{t} \left[2 a_{+} a^{*}_{-} - \left(a^{*}_{+} a_{+} + a^{*}_{-} a_{-}\right)\right].
\end{align}
Here, $\mathcal{C}$ denotes integration along the Schwinger--Keldysh closed-time-path contour, $T_{\mathbf{ij}}$ is the nearest-neighbor matrix, and the standard Einstein summation convention over the lattice indices $\mathbf{i},\mathbf{j} \in \lbrace 1,\ldots,N\rbrace$ and the replica indices $\sigma,\sigma' \in \lbrace 1,\ldots,N_s\rbrace$ is implied. The Rydberg interaction term can be decoupled using the Hubbard--Stratonovich (HS) transformation to wit
\begin{equation}
	\label{eq:Sint}
	S_{\mathrm{int}}
	=
	\int_{\mathcal{C}}\dd{t}\left[
	\im \tilde{g}\phi \eta_{\mathbf{i},\sigma}^y \eta_{\mathbf{i},\sigma}^z
	+
	\frac{1}{2} \chi_{\mathbf{i}} V^{-1}_{\mathbf{ij}} \chi_{\mathbf{j}} + \im \eta_{\mathbf{i},\sigma}^x\eta_{\mathbf{i},\sigma}^y\chi_{\mathbf{i}}\right],
\end{equation}
where we have introduced the notations $V_{\mathbf{ij}} = (\lambda/2N_s) T_{\mathbf{ij}}$, $\tilde{g} = g\sqrt{8\omega_0/N N_s}$, and a new parametrization for the photon field, $\hat{a} = \sqrt{\omega_0/2}\left(\hat{\phi} + \im\hat{\pi}/\omega_0\right)$. Performing the Hubbard–Stratonovich transformation offers two key advantages. First, it enables to efficiently resum infinitely many diagrams at a finite loop order. For example, the diagram (b) shown below corresponds to resumming an infinite series of ``bubble-chain'' diagrams \cite{Aarts:2002dj}. Second, as discussed later in the Supplementary Material, the correlation functions of the auxiliary field $\chi$ can be directly mapped to the spin-$z$ (in our case) correlation functions.

The 2PI effective action can be written in the standard form 
\begin{equation}
	\label{eq:Gamma2PI}
	\Gamma_{2\mathrm{PI}}
	=
	S + \frac{\im}{2}\Tr\ln D^{-1} + \frac{\im}{2}\Tr D_0^{-1} D - \frac{\im}{2} \Tr\ln G^{-1} - \frac{\im}{2}\Tr G_0^{-1}G + \Gamma_2\,,
\end{equation}
where $\Gamma_2 = -\im \ln\left\langle\exp\left(\im S_{\mathrm{int}}\right)\right\rangle_{\mathrm{2PI}}$ is the sum of all two-particle irreducible, with respect to the full propagators $D$ and $G$, connected vacuum diagrams and
\begin{equation}
	\label{eq:D0_G0}
	\left(\im D^{-1}_0\right)^{ab}(t,t')
	=
	\frac{\delta^2 S[\eta,\psi,\chi]}{\delta\psi^a(t)\,\delta\psi^b(t')}\,,\quad
	\left(\im D^{-1}_0\right)_{\mathbf{ij}}^{\chi\chi}(t,t')
	=
	\frac{\delta^2 S[\eta,\psi,\chi]}{\delta\chi_{\mathbf{i}}(t)\,\delta\chi_{\mathbf{j}}(t')}\,,\quad
	\left(\im G^{-1}_0\right)^{\alpha\beta}_{\mathbf{ij},\sigma\sigma'}(t,t')
	=
	\frac{\overrightarrow{\delta}}{\delta\eta^{\alpha}_{\mathbf{i},\sigma}(t)} S[\eta,\psi,\chi]\frac{\overleftarrow{\delta}}{\delta\eta^{\beta}_{\mathbf{j},\sigma'}(t')}\,,
\end{equation}
with $\boldsymbol{\psi} = (\phi,\pi)^T$, are the classical inverse propagators for the photon, the Hubbard--Stratonovich, and the fermion fields, respectively. Introducing the diagrammatic notation
\begin{align*}
	\tilde{g}\int_{\mathcal{C}}\dd{t}    
	&\ = \
	\begin{tikzpicture}[anchor=base, baseline=-0.1cm]
		\filldraw (-4ex,0) circle (0.4ex);
		\draw (-6ex,3ex) -- (-4ex,0);
		\draw (-6ex,-3ex) -- (-4ex,0);
		\draw[style=photon] (-4ex,0) to (2ex,0);
	\end{tikzpicture}
	\hspace{5ex}
	D^{\phi\phi}(t,t')
	\ = \
	\begin{tikzpicture}
		\draw[style=photon] (-4ex,0) to (4ex,0);
	\end{tikzpicture}
	\hspace{5ex}
	G(t,t')
	\ = \
	\begin{tikzpicture}[anchor=base, baseline=-0.1cm]
		\draw (-4ex,0) to (4ex,0);
	\end{tikzpicture}\\
	\int_{\mathcal{C}}\dd{t}    
	&\ = \
	\begin{tikzpicture}[anchor=base, baseline=-0.1cm]
		\filldraw (-4ex,0) circle (0.4ex);
		\draw (-6ex,3ex) -- (-4ex,0);
		\draw (-6ex,-3ex) -- (-4ex,0);
		\draw[dashed] (-4ex,0) to (2ex,0);
	\end{tikzpicture}
	\hspace{5ex}
	D^{\chi\chi}(t,t')
	\ = \
	\begin{tikzpicture}[anchor=base, baseline=-0.1cm]
		\draw[dashed] (-4ex,0) to (4ex,0);
	\end{tikzpicture}
	\hspace{5ex}
	D^{\phi\chi}(t,t')
	\ = \
	\begin{tikzpicture}[anchor=base, baseline=-0.1cm]
		\draw[style=photon] (-4ex,0) to (0,0);
		\draw[dashed] (0,0) to (4ex,0);
	\end{tikzpicture}
\end{align*}
the (next-to-)leading-order diagrams in $1/N$ and $1/N_s$ read, as will be shown below,
\begin{center}
	\begin{tikzpicture}
		\node at (-8ex,4ex) {(a)};
		\filldraw (-4ex,0) circle (0.5ex);
		\filldraw (4ex,0) circle (0.5ex);
		\draw (0,0) circle (4ex);
		\draw[style=photon] (-4ex,0) to (4ex,0);
	\end{tikzpicture}
	\hspace{5ex}
	\begin{tikzpicture}
		\node at (-8ex,4ex) {(b)};
		\filldraw (-4ex,0) circle (0.5ex);
		\filldraw (4ex,0) circle (0.5ex);
		\draw (0,0) circle (4ex);
		\draw[dashed] (-4ex,0) to (4ex,0);
	\end{tikzpicture} 
	\hspace{5ex}
	\begin{tikzpicture}
		\node at (-8ex,4ex) {(c)};
		\filldraw (-4ex,0) circle (0.5ex);
		\filldraw (4ex,0) circle (0.5ex);
		\draw (0,0) circle (4ex);
		\draw[style=photon] (-4ex,0) to (0,0);
		\draw[dashed] (0,0) to (4ex,0);
	\end{tikzpicture}  
	\hspace{5ex}
	\begin{tikzpicture}
		\filldraw (-4ex,0) circle (0.5ex);
		\filldraw (4ex,0) circle (0.5ex);
		\draw (0,0) circle (4ex);
		\draw[dashed] (-4ex,0) to (0,0);
		\draw[style=photon] (0,0) to (4ex,0);
	\end{tikzpicture}  
\end{center}
Analytically, they correspond to 
\begin{align}
	\Gamma_2^{(a)} 
	&=
	-\frac{\im \tilde{g}^2}{2} \sum_{\mathbf{i},\mathbf{j}} \sum_{\sigma,\sigma'} \int_{\mathcal{C}}\dd{t}\dd{t'} D^{\phi\phi}(t,t') \left[G_{\mathbf{ij},\sigma\sigma'}^{yz}(t,t')\,G^{zy}_{\mathbf{ij},\sigma\sigma'}(t,t') - G_{\mathbf{ij},\sigma\sigma'}^{yy}(t,t')\,G^{zz}_{\mathbf{ij},\sigma\sigma'}(t,t')\right],\nonumber\\
	\Gamma_2^{(b)} 
	&=
	-\frac{\im}{2} \sum_{\mathbf{i},\mathbf{j}} \sum_{\sigma,\sigma'} \int_{\mathcal{C}}\dd{t}\dd{t'} D^{\chi\chi}_{\mathbf{ij}}(t,t') \left[G_{\mathbf{ij},\sigma\sigma'}^{xy}(t,t')\,G^{yx}_{\mathbf{ij},\sigma\sigma'}(t,t') - G_{\mathbf{ij},\sigma\sigma'}^{xx}(t,t')\,G^{yy}_{\mathbf{ij},\sigma\sigma'}(t,t')\right],\nonumber\\
	\Gamma_2^{(c)} 
	&=
	-\frac{\im\tilde{g}}{2} \sum_{\mathbf{i},\mathbf{j}} \sum_{\sigma,\sigma'} \int_{\mathcal{C}}\dd{t}\dd{t'} \left\lbrace D^{\phi\chi}_{\mathbf{i}}(t,t') \left[G_{\mathbf{ij},\sigma\sigma'}^{yy}(t,t')\,G^{zx}_{\mathbf{ij},\sigma\sigma'}(t,t') - G_{\mathbf{ij},\sigma\sigma'}^{yx}(t,t')\,G^{zy}_{\mathbf{ij},\sigma\sigma'}(t,t')\right]\right.
	\nonumber\\
	&\phantom{-\frac{\im\tilde{g}}{2} \sum_{\mathbf{i},\mathbf{j}} \sum_{\sigma,\sigma'} \int_{\mathcal{C}}\dd{t}\dd{t'} }
	+
	\left. D^{\chi\phi}_{\mathbf{i}}(t,t') \left[G_{\mathbf{ij},\sigma\sigma'}^{yy}(t,t')\,G^{xz}_{\mathbf{ij},\sigma\sigma'}(t,t') - G_{\mathbf{ij},\sigma\sigma'}^{xy}(t,t')\,G^{yz}_{\mathbf{ij},\sigma\sigma'}(t,t')\right]\right\rbrace.
\end{align}

The action \eqref{eq:S} is invariant under $\mathbb{Z}_2$ gauge transformations $\boldsymbol{\eta}_{\mathbf{i},\sigma}(t) \to -\boldsymbol{\eta}_{\mathbf{i},\sigma}(t)$. This symmetry reflects the artificial nature of the Majorana fermions representing the physical degrees of freedom (spins), so any physical initial state must also respect it. The correlation functions accordingly transform as $G_{\mathbf{ij},\sigma\sigma'}^{\alpha\beta}(t,t') \to (-1)^{\zeta(\mathbf{i},\sigma) + \zeta(\mathbf{j},\sigma')} G_{\mathbf{ij}}^{\alpha\beta}(t,t')$, where $\zeta\in \lbrace 0, 1\rbrace$. Therefore, only diagonal entries are gauge-invariant, implying that $G_{\mathbf{ij},\sigma\sigma'}^{\alpha\beta}(t,t') = G_{\mathbf{i},\sigma}^{\alpha\beta}(t,t')\,\delta_{\mathbf{ij}}\,\delta_{\sigma\sigma'}$. With this, the proper self-energies,
\begin{equation}
	\Sigma^{\phi\phi}(t,t') \equiv 2\im \frac{\delta\Gamma_2[D,G]}{\delta D^{\phi\phi}(t',t)}\,,\quad 
	\Sigma^{\chi\chi}_{\mathbf{ij}}(t,t') \equiv 2\im \frac{\delta\Gamma_2[D,G]}{\delta D_{\mathbf{ji}}^{\phi\phi}(t',t)}\,,\quad
	\Sigma^{\phi\chi}_{\mathbf{i}}(t,t') \equiv 2\im \frac{\delta\Gamma_2[D,G]}{\delta D_{\mathbf{i}}^{\chi\phi}(t',t)}\,,\quad
	\Pi_{\mathbf{ij},\sigma\sigma'}^{\alpha\beta}(t,t') \equiv -2\im\frac{\delta\Gamma_2[D,G]}{\delta G_{\mathbf{ji},\sigma'\sigma}^{\beta\alpha}(t',t)}\,,
\end{equation}
are given by
\begin{align}
	\label{eq:self_energies}
	\Sigma^{\phi\phi}
	&=
	\tilde{g}^2 \sum_{\mathbf{i},\sigma}\left(G_{\mathbf{i},\sigma}^{yz} G_{\mathbf{i},\sigma}^{zy} - G_{\mathbf{i},\sigma}^{yy} G_{\mathbf{i},\sigma}^{zz}  \right),\quad
	\Sigma_{\mathbf{ij}}^{\chi\chi}
	=
	\sum_{\sigma}
	\left(G^{xy}_{\mathbf{i},\sigma} G^{yx}_{\mathbf{i},\sigma} - G^{xx}_{\mathbf{i},\sigma} G^{yy}_{\mathbf{i},\sigma}\right) \delta_{\mathbf{ij}}\equiv \Sigma^{\chi\chi}_{\mathbf{i}}\,\delta_{\mathbf{ij}}\,,\quad
	\Sigma^{\phi\chi}_{\mathbf{i}}
	=
	\tilde{g} \sum_{\sigma} \left(G^{yy}_{\mathbf{i},\sigma} G^{zx}_{\mathbf{i},\sigma} - G^{yx}_{\mathbf{i},\sigma} G^{zy}_{\mathbf{i},\sigma}\right),\nonumber\\
	\Pi_{\mathbf{ij},\sigma\sigma'}^{\alpha\beta} 
	&= 
	\begin{pmatrix}
		-D^{\chi\chi}_{\mathbf{ii}}G_{\mathbf{i},\sigma}^{yy} & D^{\chi\chi}_{\mathbf{ii}}G_{\mathbf{i},\sigma}^{yx} - \tilde{g} D^{\chi\phi}_{\mathbf{i}} G_{\mathbf{i},\sigma}^{yz} & \tilde{g} D^{\chi\phi}_{\mathbf{i}} G_{\mathbf{i},\sigma}^{yy} \\
		D^{\chi\chi}_{\mathbf{ii}}G_{\mathbf{i},\sigma}^{xy} - \tilde{g} D^{\phi\chi}_{\mathbf{i}} G_{\mathbf{i},\sigma}^{zy} & -D^{\chi\chi}_{\mathbf{ii}} G_{\mathbf{i},\sigma}^{xx} - \tilde{g}^2 D^{\phi\phi} G_{\mathbf{i},\sigma}^{zz} + \tilde{g}\left(D^{\chi\phi}_{\mathbf{i}} G_{\mathbf{i},\sigma}^{xz} + D^{\phi\chi}_{\mathbf{i}} G_{\mathbf{i},\sigma}^{zx}\right) & \tilde{g}^2 D^{\phi\phi}G_{\mathbf{i},\sigma}^{zy} - \tilde{g} D^{\chi\phi}_{\mathbf{i}}G^{xy}_{\mathbf{i},\sigma} \\
		\tilde{g} D^{\phi\chi}_{\mathbf{i}} G_{\mathbf{i},\sigma}^{yy} & \tilde{g}^2 D^{\phi\phi} G_{\mathbf{i},\sigma}^{yz} - \tilde{g} D^{\phi\chi}_{\mathbf{i}}G^{yx}_{\mathbf{i},\sigma} & -\tilde{g}^2 D^{\phi\phi} G_{\mathbf{i},\sigma}^{yy}
	\end{pmatrix}\delta_{\mathbf{ij}}\delta_{\sigma\sigma'}
	\equiv 
	\Pi_{\mathbf{i},\sigma}^{\alpha\beta}\,\delta_{\mathbf{ij}}\delta_{\sigma\sigma'}\,,
\end{align}
with all the temporal arguments above assumed to be $(t,t')$.

A simple power counting reveals $\Gamma_2^{(a)} = O(N^0,N_s^0)$, as one would expect. The remaining diagrams involve $D_{\mathbf{ii}}^{\chi\chi}$ and $D_{\mathbf{i}}^{\chi\phi}$, which are both zero at the bare level. To estimate how these contributions scale with $N$ and $N_s$, one can then use perturbation theory. The Dyson equations for $D_{\mathbf{ij}}^{\chi\chi}$ and $D_{\mathbf{i}}^{\chi\phi}$ read
\begin{align}
	\label{eq:DysonHS}
	D^{\chi\chi}_{\mathbf{ij}}(t,t')
	&=
	\im V_{\mathbf{ij}}\delta_{\mathcal{C}}(t-t') + \im V_{\mathbf{im}}\int_{\mathcal{C}}\dd{t''} \left[\Sigma_{\mathbf{ml}}^{\chi\chi}(t,t'')\,D^{\chi\chi}_{\mathbf{lj}}(t'',t') + \Sigma^{\chi\phi}_{\mathbf{m}}(t,t'')\,D^{\phi\chi}_{\mathbf{j}}(t'',t')\right],\nonumber\\
	D^{\chi\phi}_{\mathbf{i}}(t,t')
	&=
	\im V_{\mathbf{im}}\int_{\mathcal{C}}\dd{t''}\left[\Sigma^{\chi\phi}_{\mathbf{m}}(t,t'')\,D^{\phi\phi}(t'',t') + \Sigma^{\chi\chi}_{\mathbf{mj}}(t,t'')\,D_{\mathbf{j}}^{\chi\phi}(t'',t') \right].
\end{align}
To zeroth order, $D_{\mathbf{ij}}^{(0),\chi\chi}(t,t') = \im V_{\mathbf{ij}}\,\delta_{\mathcal{C}}(t-t')$ and $D_{\mathbf{i}}^{(0),\phi\chi}(t,t') = 0$, which yields $D^{(1),\chi\chi}_{\mathbf{ii}}(t,t') = -\sum_{\mathbf{m}} V_{\mathbf{im}} \,\Sigma^{\chi\chi}_{\mathbf{m}}(t,t') V_{\mathbf{mi}}$ and $D^{(1),\chi\phi}_{\mathbf{i}}(t,t') = \im \sum_{\mathbf{m}} V_{\mathbf{im}}\int_{\mathcal{C}}\dd{t''} \Sigma^{\chi\phi}_{\mathbf{m}}(t,t'')\,D^{\phi\phi}(t'',t')$. When summing over $\mathbf{i}$ in $\Gamma_2^{(b)}$ with $D^{(1),\chi\chi}_{\mathbf{ii}}$ the nearest-neighbor operator squared will give a factor of $z N$. The Hubbard--Stratonovich self-energy scales as $\Sigma^{\chi\chi}_{\mathbf{ij}} = O(N^0,N_s^1)$, which together with the sum over $\sigma$ in $\Gamma_2^{(b)}$ gives $N_s^2$, canceling the $1/N_s^2$ prefactor from $V^2$, resulting in $\Gamma^{(b)}_2 = O(N^1,N_s^0)$. As anticipated, the correction stemming from the short-range Rydberg interaction is not suppressed by the system size $N$. Finally, since $\Sigma^{\chi\phi}_{\mathbf{i}} = O(N^{-1/2},N_s^{1/2})$ and the sum over $\mathbf{m}$ in $D^{(1),\chi\phi}_{\mathbf{i}}(t,t')$ goes only over the nearest neighbors, we conclude $D^{(1),\chi\phi}_{\mathbf{i}}(t,t') = O(N^{-1/2},N_s^{-1/2})$, with the additional $1/N_s$ factor coming from $V$. Combined with the additional $\tilde{g} = O(N^{-1/2},N_s^{-1/2})$ and the sum over the diagonal entries $\sum_{\mathbf{i},\sigma}$ in $\Gamma_2^{(c)}$, one finds $\Gamma_2^{(c)} = O(N^0,N_s^0)$. Therefore, each considered diagram is suppressed by either $1/N$, $1/N_s$, or both. 

The nonequilibirum Dyson equations for the two-point functions $D$ and $G$ can be obtained by extremizing the 2PI effective action \eqref{eq:Gamma2PI} with respect to the propagators, $\delta\Gamma_{\mathrm{2PI}}/\delta D = 0$ and $\delta\Gamma_{\mathrm{2PI}}/\delta G = 0$, yielding
\begin{equation}
	\label{eq:KB}
	\left[\left(D_0^{-1} - \Sigma \right) \circ D\right]_{\mathrm{ab}}(t,t') =
	\delta_{\mathrm{ab}}\,\delta_{\mathcal{C}}(t-t')\,,\quad
	\left[\left(G_0^{-1} - \Pi \right) \circ G\right]^{\alpha\beta}_{\sigma\sigma',\mathbf{ij}}(t,t')  
	=
	\delta^{\alpha\beta}\,\delta_{\mathbf{ij}}\,\delta_{\sigma\sigma'}\delta_{\mathcal{C}}(t-t')\,,
\end{equation}
where $\circ$ denotes a generalized convolution that sums and integrates over all possible indices and coordinates. Inserting then the self-energies \eqref{eq:self_energies}, together with the bare inverse propagators \eqref{eq:D0_G0}, into the Dyson equations results in a closed system of equations. For example, using 
\begin{equation}
	\left(\im G^{-1}_0\right)^{\alpha\beta}_{\mathbf{ij},\sigma\sigma'}(t,t')
	=
	\frac{\overrightarrow{\delta}}{\delta\eta^{\alpha}_{\mathbf{i},\sigma}(t)} S[\eta,\psi,\chi]\frac{\overleftarrow{\delta}}{\delta\eta^{\beta}_{\mathbf{j},\sigma'}(t')} 
	= 
	\delta_{\mathcal{C}}(t-t')\,\delta_{\mathbf{ij}}\,\delta_{\sigma\sigma'}\,\left[\im \delta^{\alpha\beta}\,\partial_{t'} - M^{\alpha\beta}(t)\right],\quad 
	\im M(t)
	=
	\begin{pmatrix}
		0 & \Delta'(t) & 0 \\
		-\Delta'(t) & 0 & \tilde{g}\phi(t) \\ 
		0 & -\tilde{g}\phi(t) & 0
	\end{pmatrix},   
\end{equation}
where $\phi(t) = \langle\hat{\phi}(t)\rangle$ and $\Delta'(t) = \Delta + \lambda + \langle\hat{\chi}_{\mathbf{i}}(t)\rangle =  \Delta + \lambda + \sum_{\mathbf{j},\sigma} V_{\mathbf{ij}} \langle\hat{s}^z_{\mathbf{j},\sigma}(t)\rangle$, one readily obtains the equation for the contour Majorana propagator:
\begin{equation}
	\left[\delta^{\alpha\gamma}\partial_t + \im M^{\alpha\gamma}(t)\right]G^{\gamma\beta}(t,t',\mathbf{i},\sigma) - \int_{\mathcal{C}}\dd{t''}\Pi^{\alpha\gamma}(t,t'',\mathbf{i},\sigma)\,G^{\gamma\beta}_f(t'',t',\mathbf{i},\sigma) = \delta^{\alpha\beta}\,\delta_{\mathbf{ij}}\,\delta_{\sigma\sigma'}\delta_{\mathcal{C}}(t-t')\,.
\end{equation}
This equation can be readily transformed into a system of equations for the statistical and spectral functions by using the decomposition \cite{Berges:2004yj}
\begin{equation}
	\label{eq:prop_decomp}
	G(t,t') = F_f(t,t') - \frac{\im}{2} \rho_f(t,t')\,\mathrm{sgn}_{\mathcal{C}}(t - t')\,,\quad
	\Pi(t,t') = -\im\Pi^{(0)}(t)\,\delta_{\mathcal{C}}(t-t') + \Pi_F(t,t') - \frac{\im}{2}\Pi_{\rho}(t,t')\,\mathrm{sgn}_{\mathcal{C}}(t - t')
\end{equation}
to yield
\begin{align}
	\label{eq:KB_majorana}
	\left[\delta^{\alpha\gamma}\partial_t + \im M^{\alpha\gamma}(t)\right]\rho_f^{\gamma\beta}(t,t',\mathbf{i},\sigma) 
	&=
	-\im\int_{t'}^t\dd{t''}\Pi^{\alpha\gamma}_{\rho}(t,t'',\mathbf{i},\sigma)\,\rho^{\gamma\beta}_f(t'',t',\mathbf{i},\sigma)\,,\nonumber\\
	\left[\delta^{\alpha\gamma}\partial_t + \im M^{\alpha\gamma}(t)\right]F_f^{\gamma\beta}(t,t',\mathbf{i},\sigma) 
	&=
	-\im\int_{t_0}^t\dd{t''}\Pi^{\alpha\gamma}_{\rho}(t,t'',\mathbf{i},\sigma)\,F^{\gamma\beta}_f(t'',t',\mathbf{i},\sigma) +
	\im\int_{t_0}^{t'}\dd{t''}\Pi^{\alpha\gamma}_F(t,t'',\mathbf{i},\sigma)\,\rho_f^{\gamma\beta}(t'',t',\mathbf{i},\sigma)\,.
\end{align}
Note that $\Pi^{(0)} = 0$ in our case. In this form, the equations are often referred to as the Kadanoff--Baym equations. As a side remark, we also note that, in contrast to bosons, the fermionic statistical and spectral functions correspond to the expectation values of the commutator and anti-commutator, respectively. The fermionic version of the fluctuation-dissipation relation can be written analogously to the bosonic one \eqref{eq:FDR}, with $n_{\mathrm{eq}}(\omega)$ replaced by $-n_{\mathrm{eq}}(\omega)$, and $\coth{(\omega/2T)}$ by $\tanh{(\omega/2T)}$.

The boson propagator equations can be derived in a similar fashion. The explicit matrix form of equation \eqref{eq:KB}, in our case, reads
\begin{equation}
	\begin{pmatrix}
		\left(D_0^{-1}\right)^{\phi\phi} - \Sigma^{\phi\phi} & \left(D_0^{-1}\right)^{\phi\pi} & -\Sigma^{\phi\chi} \\
		\left(D_0^{-1}\right)^{\pi\phi} & \left(D_0^{-1}\right)^{\pi\pi} & 0 \\
		-\Sigma^{\chi\phi} & 0 & \left(D_0^{-1}\right)^{\chi\chi} -  \Sigma^{\chi\chi}
	\end{pmatrix}
	\circ
	\begin{pmatrix}
		\vphantom{\left(D_0^{-1}\right)^{\phi\phi}} D^{\phi\phi} & D^{\phi\pi} & D^{\phi\chi} \\
		\vphantom{\left(D_0^{-1}\right)^{\phi\phi}} D^{\pi\phi} & D^{\pi\phi} & D^{\pi\chi} \\
		\vphantom{\left(D_0^{-1}\right)^{\phi\phi}} D^{\chi\phi} & D^{\chi\pi} & D^{\chi\chi}
	\end{pmatrix}
	=
	\begin{pmatrix}
		\vphantom{\left(D_0^{-1}\right)^{\phi\phi}} \mathds{1} & 0 & 0 \\
		\vphantom{\left(D_0^{-1}\right)^{\phi\phi}} 0 & \mathds{1} & 0 \\
		\vphantom{\left(D_0^{-1}\right)^{\phi\phi}} 0 & 0 & \mathds{1}
	\end{pmatrix}\,,
\end{equation}
with the self-energies \eqref{eq:self_energies} and the bare inverse propagators obtained by differentiating the Keldysh action according to Eq.~\eqref{eq:D0_G0}. Two of the resulting equations are given by \eqref{eq:DysonHS}. We note that the equation for $D^{\chi\chi}$, without the last term, has the structure of the Bethe--Salpeter equation, reflecting the composite nature of the Hubbard--Stratonovich field $\chi$. The remaining equations, governing the dynamics of the photon degrees of freedom, have an integro-differential structure similar to \eqref{eq:KB_majorana} after decomposition \eqref{eq:prop_decomp}. Due to the photon-loss dissipative term, which couples the two branches of the Schwinger--Keldysh contour to each other, cf. Eq.~\eqref{eq:S}, the explicit form of the equations cannot be so easily written in a compact fashion. For the sake of completeness, we provide the Kadanoff--Baym equations for the $\pi\pi$-component of the photon propagator as an example:
\begin{align}
	\partial_t\rho^{\pi\pi}(t,t') 
	&=
	-\omega_0^2\,\rho^{\phi\pi}(t,t') - \kappa\rho^{\pi\pi}(t,t')\,\mathrm{sgn}(t-t') -
	\int_{t'}^t\dd{t''}\left[\Sigma^{\phi\phi}_{\rho}(t,t'')\,\rho^{\phi\pi}(t'',t') + \Sigma^{\phi\chi}_{\rho,\mathbf{i}}(t,t'')\,\rho_{\mathbf{i}}^{\chi\pi}(t'',t')\right],\nonumber\\
	\partial_t F^{\pi\pi}(t,t') 
	&=
	-\omega_0^2\,F^{\phi\pi}(t,t') - \kappa F^{\pi\pi}(t,t') - \kappa \omega_0 \rho^{\phi\pi}(t,t')\,\Theta(t'-t) -
	\int_{t_0}^t\dd{t''}\left[\Sigma^{\phi\phi}_{\rho}(t,t'')\,F^{\phi\pi}(t'',t') + \Sigma^{\phi\chi}_{\rho,\mathbf{i}}(t,t'')\,F_{\mathbf{i}}^{\chi\pi}(t'',t')\right]\nonumber\\
	&+
	\int_{t_0}^{t'}\dd{t''} \left[\Sigma^{\phi\phi}_F(t,t'')\,\rho^{\phi\pi}(t'',t') + \Sigma_{F,\mathbf{i}}^{\phi\chi}(t,t'')\,\rho_{\mathbf{i}}^{\chi\pi}(t'',t')\right].
\end{align}

Finally, the Kadanoff--Baym equations are accompanied by evolution equations for the one-point functions. Due to the aforementioned $\mathbb{Z}_2$ gauge symmetry, the Majorana expectation value vanishes, $\langle\hat{\boldsymbol{\eta}}\rangle = 0$. The equation for the expectation value of the Hubbard--Stratonovich field can be obtained by varying the 2PI effective action with respect to $\chi$:
\begin{equation}
	\frac{\delta \Gamma_{\mathrm{2PI}}}{\delta\chi_{\mathbf{i}}(t)} = \frac{\delta S[\eta,\psi,\chi]}{\delta \chi_{\mathbf{i}}(t)}
	-
	\frac{\im}{2}\frac{\delta \Tr\left[G_0^{-1}(\psi,\chi)\,G\right]}{\delta \chi_{\mathbf{i}}(t)} 
	=
	0 
	\; \implies \;
	\chi_{\mathbf{i}}(t) = \sum_{\mathbf{j},\sigma}\frac{\im V_{\mathbf{ij}}}{2} \left[G^{yx}(t,t,\mathbf{j},\sigma) - G^{xy}(t,t,\mathbf{j},\sigma)\right]
	=
	\sum_{\mathbf{j},\sigma} V_{\mathbf{ij}} s^z_{\mathbf{j},\sigma}(t)\,.
\end{equation}
Similarly, the equations for the photon one-point functions can be obtained by varying $\Gamma_{\mathrm{2PI}}$ with respect to $\phi$ and $\pi$, giving
\begin{equation}
	\label{eq:1pt_eq}
	\partial_t \phi 
	=
	\pi - \kappa\phi\,,\quad
	\partial_t\pi 
	=
	-\omega_0^2\phi - \kappa\pi - \tilde{g} \sum_{\mathbf{i},\sigma} s^x_{\mathbf{i},\sigma}
	=
	-\omega_0^2\phi - \kappa\pi + \im \tilde{g} \sum_{\mathbf{i},\sigma} G^{yz}(t,t,\mathbf{i},\sigma)\,.
\end{equation}
Above, we adopted a simplified notation $\chi = \langle\hat{\chi}\rangle$, $\phi = \langle\hat{\phi}\rangle$, etc.

\section{Details of the numerical implementation}
Numerical approaches to solving the Kadanoff--Baym equations are covered in great detail in, e.g., \cite{Schuckert:2018fwg,Meirinhos:2022,Hosseinabadi:2023cyj}. Mathematically, the problem amounts to solving a set of coupled nonlinear Volterra integro-differential equations:
\begin{equation}
	y_i'(t) = \Phi_i[t, \mathbf{y}(t)] + \int\limits_{\mathcal{D} \in [t_0,t]}\dd{s} K_i[t, s, \mathbf{y}(s)] \equiv g_i(t,\mathbf{y})\,,\quad \mathbf{y}(t_0) = \mathbf{y}_0.
\end{equation}
Note that the Bethe--Salpeter-type equations \eqref{eq:DysonHS} have a similar form, albeit with $y_i'(t) \equiv 0$ on the left-hand side. We solve these equations numerically using the iterative Heun's scheme,

\begin{align}
	\mathrm{predictor:}\quad\mathbf{y}_{n+1}^{(0)} = \mathbf{y}_n + \Delta t\,\mathbf{g}(t_n,\mathbf{y}_n)\,,\quad \mathrm{corrector:}\quad
	\mathbf{y}_{n+1}^{(k+1)} = \mathbf{y}_n + \frac{\Delta t}{2}\left[\mathbf{g}(t_n,\mathbf{y}_n) + \mathbf{g}(t_{n+1},\mathbf{y}^{(k)}_{n+1})\right],
\end{align}
and employing the trapezoidal rule to compute the memory integrals. The correction step is iterated until the desired convergence is reached. A measure of convergence for given tolerances is 
\begin{equation}
	\epsilon^{(k)}_{n+1} = \frac{\norm{\mathbf{y}_{n+1}^{(k+1)} - \mathbf{y}_{n+1}^{(k)}}_p}{\mathrm{atol} + \mathrm{rtol}\cdot \norm{\mathbf{y}_{n+1}^{(k+1)}}_p},
\end{equation}
where $\norm{\cdot}_p$ denotes the standard $L^p$-norm, with $p=2$ chosen in this work. The step is accepted if $\epsilon^{(k)}_{n+1} < 1$. For all data presented, we used $\mathrm{atol} = 10^{-8}$ and $\mathrm{rtol} = 10^{-6}$. The timestep for all plots except Fig.~\ref{fig:Ts_ph}b was taken as $\Delta t = 0.25$ in units of $\omega_0$. The points in Fig.~\ref{fig:Ts_ph}b were evaluated on a coarser grid with timestep $\Delta t = 0.4$ instead. In all cases, the evolution time was fixed to $\tau_{\mathrm{evol}} = 400$, corresponding to $N_t = 1600$ timesteps for the finer lattice and $N_t = 1000$ timesteps for the coarser lattice, respectively. Note that $\Delta t$ is much smaller than any other timescale ($\omega_0^{-1}$, $\lambda^{-1}$, etc.) in the problem, which implies that all relevant physical processes were resolved.

\section{Spin correlation functions}
In this work, we restrict ourselves to spatially homogeneous states. In this case, it is suggestive to work in Fourier space:
\begin{equation}
	f_{\mathbf{k}} = \sum_{\mathbf{j}} \mathrm{e}^{\im \mathbf{k}\cdot\mathbf{j}} f_{\mathbf{j}}\,,\quad 
	f_{\mathbf{j}} = N^{-1} \sum_{\mathbf{k}} \mathrm{e}^{-\im\mathbf{k}\cdot\mathbf{j}} f_{\mathbf{k}}\,.
\end{equation}
To capture the antiferromagnetic nature of the Rydberg interaction, we then introduce even and odd sublattices, denoted by $A$ and $B$, respectively. Consequently, correlation functions involving spatial dependence will carry sublattice indices: $G_a$, $D^{\chi\chi}_{\mathbf{k},ab}$, $D^{\chi\phi}_{a}$, where $a,b \in \lbrace A, B\rbrace$. In addition, the original Brillouin zone is replaced by two magnetic (or reduced) Brillouin zones, each twice as small as the original one. For a two-dimensional square lattice with unit lattice spacing, the reciprocal basis of the reduced Brillouin zone (RBZ) is spanned by the vectors $\mathbf{G}_1 = (\pi, \pi)^T$ and $\mathbf{G}_2 = (\pi,-\pi)^T$. The map back to the original Brillouin zone is given by
\begin{equation}
	D_{\mathbf{k}}^{\chi\chi}
	=
	\frac{1}{2}
	\begin{cases}
		D^{\chi\chi}_{AA,\mathbf{k}} + \mathrm{e}^{-\im\mathbf{k}\cdot\mathbf{r}_0} D^{\chi\chi}_{AB,\mathbf{k}} + \mathrm{e}^{\im\mathbf{k}\cdot\mathbf{r}_0} D^{\chi\chi}_{BA,\mathbf{k}} + D^{\chi\chi}_{BB,\mathbf{k}},\quad &\mathbf{k} \in \text{RBZ}\,,\\
		D^{\chi\chi}_{AA,\mathbf{q}} + \mathrm{e}^{-\im\mathbf{k}\cdot\mathbf{r}_0}D^{\chi\chi}_{AB,\mathbf{q}} + \mathrm{e}^{\im\mathbf{k}\cdot\mathbf{r}_0}D^{\chi\chi}_{BA,\mathbf{q}} + D^{\chi\chi}_{BB,\mathbf{q}},\quad &\mathbf{q} \equiv \mathbf{k} - \mathbf{G} \in \text{RBZ}\,,
	\end{cases}
\end{equation}
where $\mathbf{r}_0$ is the displacement vector between the two sublattices, which for the simple square lattice with unit lattice spacing can be chosen as either $(1,0)^T$ or $(0,1)^T$.

As discussed in \cite{Schuckert:2018fwg}, the HS correlator can be readily mapped to a connected spin correlation function:
\begin{equation}
	\label{eq:HS_to_spin}
	\tilde{D}_{\mathbf{ij}}^{\chi\chi}(t,t') \equiv D_{\mathbf{ij}}^{\chi\chi}(t,t') -\im V_{\mathbf{ij}} \delta_{\mathcal{C}} (t - t')     
	=
	V_{\mathbf{im}} V_{\mathbf{jl}} \sum_{\sigma,\sigma'} \langle \mathcal{T}_{\mathcal{C}}\hat{s}_{\mathbf{m},\sigma}^z(t)\,\hat{s}_{\mathbf{l},\sigma'}^z(t') \rangle_c \equiv V_{\mathbf{im}} V_{\mathbf{jl}} \sum_{\sigma,\sigma'} C^{zz}_{\mathbf{ml},\sigma\sigma'}(t,t')\,.
\end{equation}
Going to momentum space, the expression takes a particularly simple form in the limit $N_s\to1$: $C^{zz}_{\mathbf{k}}(t,t') = \tilde{D}_{\mathbf{k}}^{\chi\chi}(t,t')/V_{\mathbf{k}}^2$. Decomposing the nonsingular part $\tilde{D}_{\mathbf{k}}^{\chi\chi}(t,t')$ of the HS propagator as shown in \eqref{eq:prop_decomp} then allows one to extract spectral and statistical properties of the spin degrees of freedom and thus probe for their thermalization using the generalized fluctuation-dissipation relation, as discussed in the main text. It is worth noting that, despite what Eq.~\eqref{eq:HS_to_spin} might naively suggest, the accessibility of the spin-$z$ correlation function did not rely on the Hubbard--Stratonovich transformation, nor did it depend on $V \neq 0$. In particular, one can derive a similar Bethe--Salpeter--type equations for other spin correlation functions $C^{\alpha\beta}$. However, since thermalization properties of the spin degrees of freedom are not anticipated to differ across different components, the spin-$z$ correlation functions were sufficient for the current work. For more details on how higher-order correlation function can be extracted within the 2PI formalism, we refer to \cite{Arrizabalaga:2004,Berges:2005,Carrington:2014}.

\section{Effective temperatures}
In order to reduce the influence of numerical errors when extracting effective temperatures (cf. discussion in the main text), we first average the correlation functions over a small time window, $\bar{F}(\tau,\omega) = T_1^{-1} \int_{\tau-T_1/2}^{\tau+T_1/2}\dd{\tau'} F(\tau',\omega)$ and  $\bar{\rho}(\tau,\omega) = T_1^{-1}\int_{\tau-T_1/2}^{\tau+T_1/2}\dd{\tau'} \rho(\tau',\omega)$, with $T_1 = 8$ chosen for the results presented in this work such that it is much smaller than any equilibration timescale. Therefore, the averaging procedure does not affect the slow dynamics of macroscopic observables (e.g., effective temperatures). At the same time $T_1/\Delta t \gg 1$, so that each bin contains a statistically significant number of points. For brevity, we omit the overbars elsewhere in the manuscript.

To extract the low-frequency effective temperatures, we then average over the appropriate frequency windows $(\omega_{\mathrm{min}},\omega_{\mathrm{max}})$ and perform one final temporal averaging in order to reduce the residual oscillations: 
\begin{equation}
	\label{eq:T_average}
	T_{\mathrm{eff}}(\tau) = \frac{T_2^{-1}}{\omega_{\mathrm{max}} - \omega_{\mathrm{min}}} \int_{\omega_{\mathrm{min}}}^{\omega_{\mathrm{max}}} \dd{\omega'} \int_{\tau-T_2/2}^{\tau+T_2/2}\dd{\tau'} T_{\mathrm{eff}}(\tau',\omega')\,.
\end{equation}
We choose $T_2 = 4$ in this work, which effectively smoothens oscillations with frequencies up to $2\pi/T_2 \sim 1.6$, covering all major frequencies in the excitation spectra, cf. Fig.~\ref{fig:observables}b. As before, however, $T_2$ is much smaller than any equilibration timescale. 

The frequency windows are chosen as $(\omega_{\mathrm{min}},\omega_{\mathrm{max}}) \approx (8\cdot10^{-3},7\cdot10^{-2})$ and $(8\cdot10^{-3},4\cdot10^{-1})$ for the photon and spin degrees of freedom, respectively. In both cases, $\omega_{\mathrm{min}}$ corresponds to the smallest available nonzero frequency mode set by the evolution time $\tau_{\mathrm{evol}} = 400$. In Fig.~\ref{fig:occupancies}a, we show the thermal fits together with the frequency windows $(\omega_{\mathrm{min}},\omega_{\mathrm{max}})$ employed to compute the effective temperatures $T_{\mathrm{ph}}$ and $T_{\mathrm{s}}$.

\begin{figure}[H]
	\centering
	\includegraphics[width=0.9\linewidth]{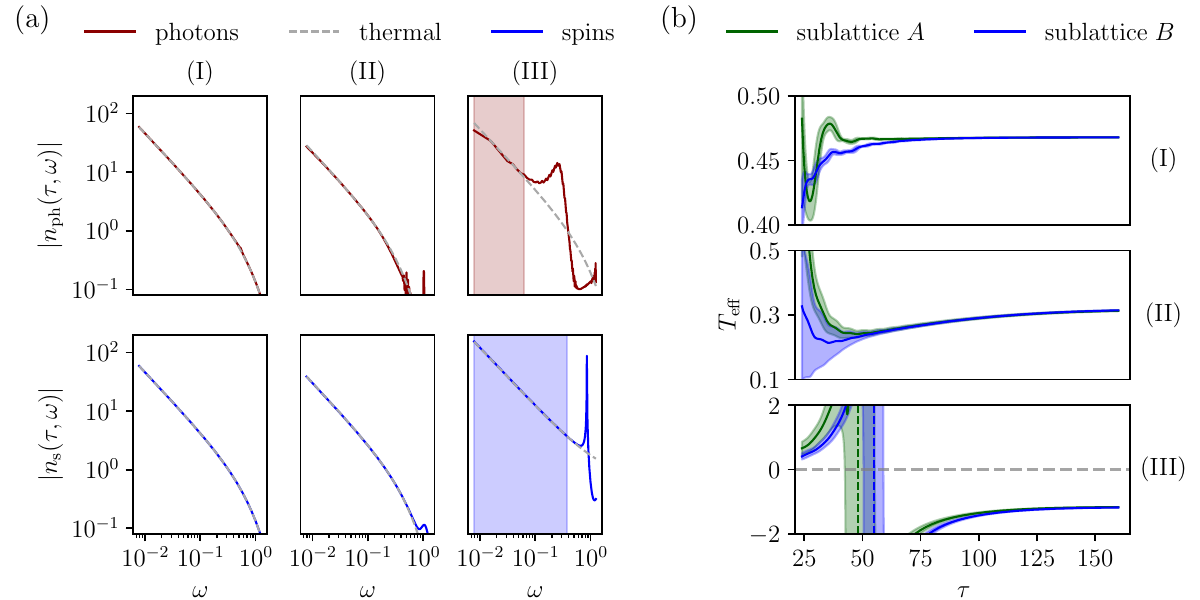}
	\caption{(a) Absolute values of the ``occupation numbers'' $n$ at $\tau = 140$ for photon and spin degrees of freedom, respectively. The shaded areas represent the frequency windows $(\omega_{\mathrm{min}},\omega_{\mathrm{max}})$ taken in Eq.~\eqref{eq:T_average} for the respective degree of freedom. For consistency, we use the same frequency windows for all the parametric regimes considered in this work. We note that, since the definition of $n$ involves the quotient of $F$ and $\rho$, cf. Eq.~\eqref{eq:FDR} and the subsequent discussion, the high-frequency region, where the value of $\rho$ drops below the numerical tolerance, suffers from the numerical artifacts and is thus not shown here. (b) Effective spin temperatures extracted from the local correlation functions on each sublattice. The two sublattices quickly thermalize with each other, exhibiting no qualitative difference throughout the entire dynamics. }
	\label{fig:occupancies}
\end{figure}

While the definition for the photon effective temperature is unambiguous, the spin effective temperature allows for more than one definition due to its rich spatial structure. In this work, to define the spin temperature, we use on-site correlation functions, which, according to the previous section, can be extracted as
\begin{equation}
	C^{zz}_{A/B}(t,t',\mathbf{r}=\mathbf{0})
	=
	\frac{2}{N}\sum_{\mathbf{k}\in \mathrm{RBZ}} \tilde{D}_{AA/BB,\mathbf{k}}^{\chi\chi}(t,t')/V_{\mathbf{k}}^2\,,
\end{equation}
where we used the fact that the number of spins on each of the sublattices is equal to half the total number of spins. For the same reason, the reduced Brillouin zones of the two sublattices are identical. Decomposing $C^{zz}_{A/B}(t,t',\mathbf{r}=\mathbf{0})$ according to \eqref{eq:prop_decomp}, we can define the effective spin temperatures on each of the sublattices. In Fig.~\ref{fig:occupancies}b, we show the local spin temperatures on each of the sublattices. As one can see, the two sublattcies quickly thermalize with each other and demonstrate no qualitative difference in their thermalization dynamics. Because of this, in all other plots shown in this work, we pick only one of them (the sublattice $B$) to present the results. 

As a final remark, we note that the spectral functions, in our convention, are given by $\rho(t,t') = \im\langle[\hat{a}(t),\hat{a}^{\dagger}(t')]\rangle$ and $\rho_s(t,t') = \im\langle[\hat{s}^z(t),\hat{s}^z(t')]\rangle$, which are conjugate antisymmetric and odd functions of the relative coordinate $t-t'$, respectively. As a result, their Wigner transforms are purely imaginary. To get real-valued Wigner transforms, we therefore slightly modify the definition as
\begin{equation}
	\rho(\tau,\omega) = -\im\int_{-2\tau}^{2\tau}\dd{s} \mathrm{e}^{\im\omega s}  \rho(\tau + s/2,\tau - s/2)\,,\quad 
	\rho_{\mathrm{s}}(\tau,\omega) = -\im\int_{-2\tau}^{2\tau}\dd{s} \mathrm{e}^{\im\omega s}  \rho_{\mathrm{s}}(\tau + s/2,\tau - s/2)\,,
\end{equation}
see also Ref.~\cite{Berges:2004yj} for a similar discussion. The statistical functions, on the other hand, are conjugate symmetric and even, respectively, and thus don't require an additional imaginary unit to make their Wigner transforms real-valued. 

\section{Photon number and superradiance}
As discussed in the main text, macroscopically large photon correlation functions, $\langle\hat{a}^2\rangle \sim N$, signal the presence of superradiance. In Fig.~\ref{fig:Nph}, we show the time dependence of the photon number per spin for the three parametric regimes considered in the main text. Note that while regimes II and III, characterized by strong Rydberg interactions, exhibit an asymptotically vanishing photon number, regime I sustains an asymptotically nonzero photon number per spin, indicating superradiance.

\begin{figure}[H]
	\centering
	\includegraphics[width=0.5\linewidth]{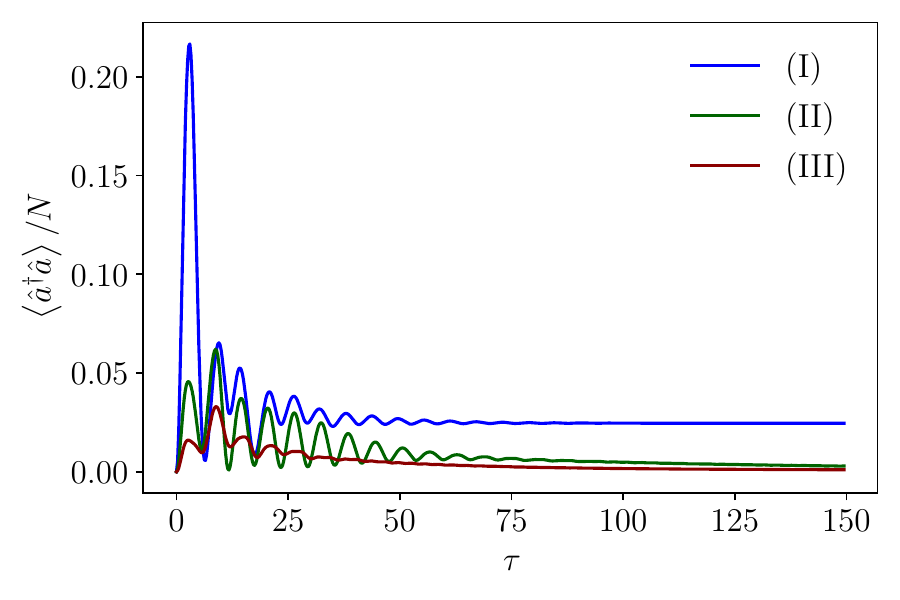}
	\caption{Number of photons per spin for the three parametric regimes discussed in the main text.}
	\label{fig:Nph}
\end{figure}
	
\end{widetext}

\bibliographystyle{apsrev4-2}
\bibliography{bibliography}

\end{document}